\documentclass[conference]{IEEEtran}

\usepackage[dvips]{graphicx}
\usepackage{amsmath,amssymb}

\newcommand{\defeq}{\stackrel{\triangle}{=}}

\newcommand{\D}{\mbox{D}}
\newcommand{\E}{\mathbb{E}}
\newcommand{\R}{\mathbb{R}}
\newcommand{\F}{\mathbb{F}}

\begin{document}

\title{Joint Quantizer Optimization based on Neural Quantizer for Sum-Product Decoder}

\author{%
  \IEEEauthorblockN{
  		Tadashi Wadayama
  		and Satoshi Takabe}
  \IEEEauthorblockA{\IEEEauthorrefmark{1}%
		Nagoya Institute of Technology,
		Gokiso, Nagoya, Aichi, 466-8555, Japan,\\
 		\{wadayama, s\_takabe\}@nitech.ac.jp}
}

\maketitle

\begin{abstract}
A low-precision analog-to-digital converter (ADC)
is required to implement a frontend device of wideband digital communication systems 
in order to reduce its power consumption.
The goal of this paper is to present a novel joint quantizer optimization method 
for minimizing lower-precision quantizers matched to the sum-product algorithms.
The principal idea is to introduce a quantizer that includes 
a feed-forward neural network and the soft staircase function. 
Since the soft staircase function is differentiable and has non-zero gradient values everywhere,
we can exploit backpropagation and a stochastic gradient descent method 
to train the feed-forward neural network in the quantizer.
The expected loss regarding the channel input and
the decoder output is minimized in a supervised training phase.
The experimental results indicate that the joint quantizer optimization method
successfully provides an 8-level quantizer for a low-density parity-check (LDPC)  code 
that achieves only a 0.1-dB performance loss compared to the unquantized system.
\end{abstract}

\IEEEpeerreviewmaketitle

\section{Introduction}

The analog-to-digital converter (ADC) is a key component 
to bridge continuous analog domain and digital domain
in digital communication systems. At the frontend of a receiver,
an ADC transforms received analog signals to digital signals 
by sampling and quantization.
Since the required bandwidth for signals is extremely broad in recent 
wideband digital communication systems,   
ADCs with high data rates are required to handle such wideband signals.
In the design phase of such a system, 
we should carefully consider the hardware cost and power consumption of ADCs
because a high precision ADC with a high data rate is known to be expensive and
very {\em power hungry} \cite{Walden, Singh}.
For example,  at the receiver side of a wideband MIMO communication system, 
a number of ADCs are needed in order to implement the analog-to-digital frontend.
Studying the potential of low-precision ADCs in MIMO detectors
with little degradation of the detection performance is important.
Mezghani et al. \cite{Mezghani} discussed combinations of 
the minimum mean squared error (MMSE) receivers for MIMO channels with 
a low-precision quantizer.
Appropriate design for quantizers matched to the MIMO detectors 
is of practical importance
because this yields a receiver with lower power consumption.

Moreover,  a design problem for a combination of a quantizer and 
a decoder for an error-correcting code (ECC) has been extensively studied.
Low-density parity-check (LDPC) codes \cite{Gallager63, MacKay99} are
a powerful class of ECCs for memoryless channels and have been practically 
exploited in a number of areas such as satellite broadcasting, 
local area wireless networks, and non-volatile storage systems.
The information bottleneck method introduced by Tishby et al. \cite{Tishby}
is becoming a common tool for designing a quantizer matched to an 
LDPC decoder \cite{Lewandowsky, Koetter, Zeitler, Kurkoski}.
Kurkoski and Yagi \cite{Kurkoski} presented 
a method to design a quantizer for binary-input discrete memoryless channels.
The quantizer designed using their method 
is optimal in the sense of maximizing mutual information 
between the channel input and the quantizer output.
Lewandowsky and  Bauch \cite{Lewandowsky} used the information bottleneck 
method to design 
discretized  message passing decoding algorithms.

Recent progress in {\em deep neural networks} (DNN) has triggered wide spread 
research activities and applications on DNNs. 
A number of practical applications 
such as image recognition, speech recognition
and robotics arise based on DNNs.
The advancement of DNNs has influences on design of algorithms 
for communications and signal processing~\cite{Com2, Com3}.

By unfolding an iterative process of an iterative signal processing algorithm, 
we can obtain a {\em signal-flow graph}, which
includes trainable variables that can be tuned using a supervised learning method, 
i.e., standard deep learning techniques such as 
stochastic gradient descent algorithms based on backpropagation and 
mini-batches can be used to adjust the trainable parameters. 
For example, Nachmani et al. \cite{Com2} presented a DNN approach to improve 
the sum-product decoder.

The joint optimization for quantizers is still a research 
topic worth studying because
{\em direct optimization} of a quantizer in terms of decoding performance has not yet been 
established. Namely, a quantizer and a decoder should be 
optimized jointly in order to minimize a given distortion measure but such an
optimization is not straightforward.
The goal of this paper is to present a novel {\em joint quantizer optimization}  method
for optimizing the quantizers matched to the sum-product algorithms.
The principal idea is to use a quantizer including 
a feed-forward neural network and the soft staircase function. 
Since the soft staircase function is differentiable 
and has non-zero gradient values everywhere, 
we can exploit  backpropagation and a stochastic gradient descent method 
to train the feed-forward neural network in 
the quantizer. In a training process, the soft staircase function 
is gradually annealed and eventually it converges 
to a discrete-valued staircase function. 
Once the training is finished, the neural quantizer can be replaced with 
a usual discrete-valued quantizer. 

\section{Preliminaries}


\subsection{Channel model}

Here, we introduce the following stochastic channel model.
Let $X$ be a random variable representing a {\em souse signal} which 
takes a value in $\mathbb{R}^n$ where $n$ is a positive integer.
The random variable $X$ follows the probability density function (PDF) $P_X(\cdot)$,
which is referred to as the prior PDF.
The channel is described using the conditional PDF $P_{Y|X}(\cdot | \cdot)$
where $Y$ is a random variable representing an observed signal 
which takes a value in  $\mathbb{R}^m$,  where $m$ is a positive integer.
The task of a {\em detector} $D(\cdot)$ is to infer a source signal 
from a received signal $Y$, namely, the detector produces the estimate signal 
$\hat X = \D(Y)$. A distortion measure 
$\mu:\mathbb{R}^n \times \mathbb{R}^n \rightarrow \mathbb{R}^n$
is used to evaluate the quality of the estimation $\hat X$. 
For example, the normalized Hamming measure 
$
\mu(X, \hat X) = (1/n) d_H(X, \hat X)
$
is often used to evaluate a detection or a decoding algorithm,
where $d_H(\cdot, \cdot)$ denotes the Hamming distance. 
A detector should be designed to minimize the {\em expected distortion}
$
 \E [\mu(X, \hat X)].	
$
Using the notation introduced here,  
a stochastic channel model can be specified with the 4-tuple: 
$(P_X, P_{Y|X}, \D, \mu)$.

\subsection{Quantizer}

A real-valued function $q: \mathbb{R} \rightarrow T$ is called a {\em quantizer function}
if  $T \subset \mathbb{R}$ 
is a finite set. A device realizing a quantization function 
is called a {\em quantizer}. For simplicity, we write 
a coordinate-wise quantization function as
$
q(x)  \defeq (q(x_1), q(x_2), \ldots, q(x_n))	
$
for $x = (x_1, x_2, \ldots, x_n) \in \mathbb{R}^n$.
The same convention is used throughout the paper.

In the present paper, we study the following scenario.
Suppose that we have a channel model $(P_X, P_{Y|X}, \D, \mu)$.
At the receiver side, we will quantize the received signal $Y$ by a quantizer $q(\cdot)$
before the detection process is carried out.
The quantized received signal $q(Y)$ is fed into the detector, and  
the estimate $\tilde X = \D(q(Y)))$ is finally obtained.

The {\em joint quantizer optimization}  discussed herein 
is the problem of designing a quantizer
to minimize the expected distortion $\E [\mu(X, \D(q(Y))) ]$.
In general, the expected distortion increases when a quantizer 
is used. In other words, we would like to find a quantizer 
without a significant increase in the expected distortion.

\subsection{Soft staircase function}

In order to tackle the joint optimization problem described above, 
we need solve an optimization problem for minimizing $\E [\mu(X, \D(q(Y))) ]$ but 
the problem is in general computationally intractable because 
it is a highly non-convex minimization problem. 
Another obstacle for this problem is that 
a quantization function is differentiable 
almost everywhere,  but
its derivative function is also zero almost everywhere.
This means that these obstacles prevent us from using backpropagation to evaluate
the gradients of the trainable variables in the quantizer.
In other words, no gradient information can pass though the quantizer 
while backward computation processes.

As a feasible approach to overcome this difficulty,  
we introduce a {\em neural quantizer}
where the corresponding quantizer function is differentiable and 
has non-zero derivative values everywhere when the 
{\em temperature parameter }  $\sigma^2$ is positive. The temperature parameter
controls the degree of smoothness of the neural quantizer function  
and the function converges to a hard 
quantizer function at the limit $\sigma^2 \rightarrow 0$.
Our strategy is to use standard deep learning techniques, such as 
backpropagation and a stochastic gradient descent method, 
to optimize the neural quantizer.
The key is to train the learnable parameters with {\em annealing}, i.e., 
the temperature parameter $\sigma^2$ is gradually decreased 
in an optimization process.
A hard quantization function that provides smaller expected distortion 
is expected to be 
obtained at the end of an optimization process.

\begin{figure}
\begin{center}
	\includegraphics[scale=0.7]{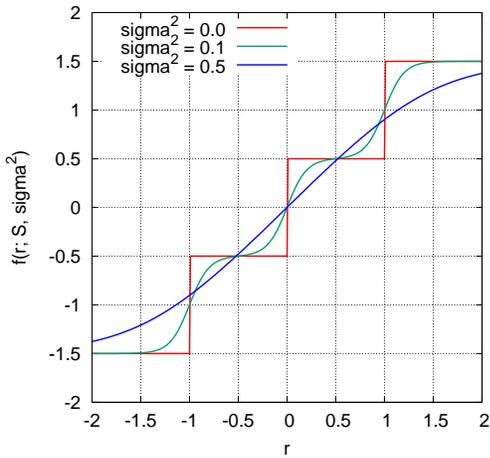}	
\end{center}
	\caption{Plot of soft staircase functions $f(r; S, \sigma^2)$ for $\sigma^2 = 0, 0.1, 0.5, S = \{-1.5, -0.5, 0.5, 1.5\}$.}
	\label{soft_q}
\end{figure}

The soft staircase function used here is the MMSE estimator function.
Let $y\in \R$ be an input to the neural quantizer.
A finite set $S \subset \R$ is called a {\em level set}.
We use the following {\em soft staircase function} $f(r; S, \sigma^2)$
as a basic component of the neural quantizer:
\begin{equation}\label{soft}
f(r; S, \sigma^2)
\defeq \frac{\sum_{s \in S} s \exp \left(- \frac{(r-s)^2}{2 \sigma^2} \right)}
{\sum_{s \in S} \exp \left(- \frac{(r-s)^2}{2 \sigma^2} \right)}, \quad \sigma^2 > 0
\end{equation}
\begin{equation} \label{solid}
f(r; S, \sigma^2)
\defeq \arg \min_{s \in S} ||r - s||_2, \quad \sigma^2 = 0,
\end{equation}
where the level set $S$ and the temperature parameter $\sigma^2$ 
controls the shape of the soft staircase.
We refer to the function in (\ref{solid}) as a {\em solid staircase function}.
Note that the soft staircase function (\ref{soft}) converges to 
the solid staircase function (\ref{solid}) at the limit $\sigma^2 \to 0$.
This justifies the definition in (\ref{solid})\footnote{For practical 
implementation, the condition $\sigma^2 > 0$ should be replaced with 
$\sigma^2 > \epsilon$ in (\ref{soft}) in order to avoid numerical instability,   
where $\epsilon$ is a small real number. In a similar manner,
the condition $\sigma^2=0$ in (\ref{solid}) should be replaced 
with $\sigma^2 \le \epsilon$.}.
Figure \ref{soft_q} presents the shapes of the soft staircase function
$f(r; S, \sigma^2)$ for $\sigma^2 = 0.0, 0.1, 0.5$ where
$S = \{-1.5, -0.5, 0.5, 1.5\}$. The curve of $\sigma^2 = 0.1$ has 
smooth wave-like shape 
but the function becomes piecewise constant 
when $\sigma^2 = 0$.

The soft staircase function 
is the MMSE estimator for a discrete signal sets over an additive white Gaussian noise (AWGN) channel.
The derivation of the MMSE estimator is briefly summarized as follows.
Let $S$ be a set of signal points:
$
	S = \{s_1, \ldots, s_M \},\ s_i \in \mathbb{R}.
$
We here assume that the prior distribution on the transmitted signal is given by
$
	p(x) = \sum_{s \in S} (1/M) \delta (x - s)
$
where $\delta (\cdot)$ is Dirac's delta function.
The received symbol  is $y = x + z$ where
$z$ represents a random noise following 
the zero mean Gaussian distribution with variance $\sigma^2$.
The conditional Probability Density Function (PDF) for this channel is thus given by
$
	p(y|x) = \frac{1}{\sqrt{2 \pi \sigma^2}} \exp \left(- {(y-x)^2}/{(2 \sigma^2)}  \right).
$

We then derive the MMSE estimator $E[x|y]$ of the original signal.
The joint PDF can be written as 
\begin{equation}
	P(x,y) =
	 \sum_{s \in S} \frac{1}{M \sqrt{2 \pi \sigma^2}} \delta(x-s) \exp \left(- \frac{(y-x)^2}{2 \sigma^2} \right).
\end{equation}
Using these quantities, the MMSE estimator $E[x|y]$ is given by
\begin{eqnarray}
	E[x|y] 	&=& \int_{-\infty}^{\infty} x p(x | y ) dx 
	= \int_{-\infty}^{\infty} \frac{x p(x, y )}{p(y)} dx \\
&=&	 \frac{\sum_{s \in S} s \exp \left(- \frac{(y-s)^2}{2 \sigma^2} \right)}{\sum_{s \in S} \exp \left(- \frac{(y-s)^2}{2 \sigma^2} \right)} = f(r; S, \sigma^2).
\end{eqnarray}

\section{Neural quantizer}
\subsection{Architecture}
The neural quantizer is a feed-forward neural network defined by 
\begin{eqnarray}
h_1 &=& relu(W_1 y + b_1 ) \\
h_i  &=& relu(W_i h_{i-1} + b_i ),\quad i \in [2, T-1]	 \\
\tilde y &=& \alpha f(W_T h_{T-1} + b_T; S, \sigma^2),
\end{eqnarray}
where $y \in \R$ is the input value and $\tilde y \in \R$  is the output value.
The function $relu$ is the ReLU function defined by $relu(x) \defeq \max\{0, x \} (x \in \R)$ and
we follow the convention $relu(x) \defeq (relu(x_1), \ldots, relu(x_n))$ for $x =(x_1, \ldots, x_n) \in \R^n$.
From some preliminary experiments, we found that the ReLU function is well behaved as an activation function
in the neural quantizer.
The vectors $h_i \in \R^{u} (i \in [1, T-1])$ are the hidden state vectors representing 
the internal states of the neural quantizer. The length of the hidden state vectors $u$ is 
called the hidden state dimension.

The trainable variables are
$W_1 \in \R^{u \times 1}, b_1 \in \R^{u}$,
$W_i \in \R^{u \times u}, b_i  \in \R^{u} \quad i \in [2, T-1]$,
$W_T \in \R^{1 \times u}, b_T \in \R$ and $\alpha \in \R$.
The set of trainable variables is compactly denoted 
by $\Theta \defeq \{W_1,\ldots, W_T, b_1,\ldots, b_T,  \alpha \}$ for simplicity.
The level set $S \defeq \{s_0, s_1, \ldots, s_{L-1}\}$ is defined by
$s_i  \defeq i - L/2 + 1/2 (i \in [0, L-1])$ for a given positive even integer $L$, 
which can be regarded as the number of quantization levels.

\begin{figure}
\begin{center}
	\includegraphics[scale=0.4]{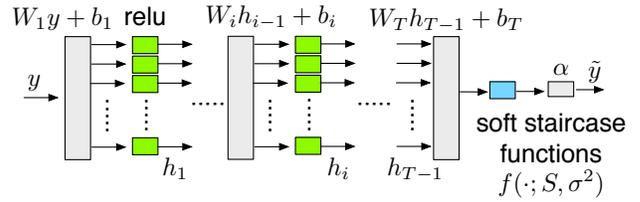}	
\end{center}
	\caption{Block diagram of the neural quantizer $\tilde y = q_{NQ}(y; \sigma^2, L)$. }
	\label{fig:nq}
\end{figure}

The entire input and output relationship of the neural quantizer is denoted by 
$q_{NQ}(\cdot; \sigma^2, L): \R \to \R$,  namely,
$
	\tilde y = q_{NQ}(y; \sigma^2, L).
$
The parameters $u$, $T$, $\sigma^2$, and $L$  are the {\em hyper parameters}
required to specify the structure of a neural quantizer.
Figure \ref{fig:nq} shows a block diagram of 
the neural quantizer defined above.

A quantization function $q: \R \to \R$ is usually used in parallel, i.e., 
$\tilde y = q(y)$ for $y \in \R^n$. In the present case,  the symbol-wise quantization 
by the neural quantizer can be described by 
$\tilde y = q_{NQ}(y; \sigma^2, L)$ for $y \in \R^n$.
This parallel quantization process can be carried out using $n$-neural quantizers.
Throughout the paper, we assume that the entire set of $n$-neural quantizers 
shares the trainable variables $\Theta$. 
This means that the $n$-neural quantizers are identical. 

\subsection{Supervised training with annealing}

In the following discussion, we focus on the squared $L_2$ distortion function 
as the distortion measure, i.e., $\mu(a, b) \defeq ||a - b||_2^2$.
Our pragmatic approach to the quantizer design problem 
is to recast the optimization problem as a minimization of the {\em expected loss}
\begin{equation}
	\E \left[\sum_{i = 1}^K ||x^{(i)}  - \D(q_{NQ}(y^{(i)}; \sigma^2, L))||_2^2 \right],
\end{equation}
where $x^{(i)}, y^{(i)} (i \in [1, K])$ are randomly generated samples of the random variables $X$ and $Y$.
The trainable variables in $\Theta$ of the neural quantizer are adjusted to lower the expected loss
by using a stochastic descent type algorithm such as SGD, Momentum, RMSprop, or Adam. If 
a detector algorithm or a decoding algorithm $\D$ is a 
differentiable function with respect to its input variable, we can use backpropagation 
to compute the gradients on the trainable variables in $\Theta$. 
The temperature parameter $\sigma^2$ should be appropriately 
decreased in a supervised training process.

The supervised training process for the neural quantizer is summarized as follows.
\begin{description}
	\item [(1)] Set $t = 1$.
	\item [(2)] Sample a mini-batch 
\[
B \defeq \{(x_1, y_1), (x_2, y_2), \ldots, (x_K, y_K) \}
\]
according to the channel model $(P_X, P_{Y|X})$, where
 $x^{(i)}, y^{(i)} (i \in [1, K])$ are realization vectors of random variables $X$ and $Y$.
	\item[(3)] After setting the objective function 
\begin{equation} \label{L2dist}
	\sum_{i = 1}^K ||x^{(i)}  - \D(q_{NQ}(y^{(i)}; \sigma^2, L))||_2^2,
\end{equation}
execute a backpropagation process to evaluate the gradient of the trainable variables in $\Theta$.
	\item[(4)] Set the temperature parameter $\sigma^2  = t^\eta$ where $\eta$ 
	is a negative constant called a {\em cooling factor}.
	\item[(5)] Update trainable variables in $\Theta$ according to the update rule 
	of a specified stochastic gradient descent algorithm. 
	\item[(6)] If $t < t_{max}$ then increment $t$ and return to Step (2). Otherwise, quit the training process.
\end{description}
Step (4) is relevant to the annealing process,  where 
the cooling factor $\eta$ controls the decay speed of the temperature parameter $\sigma^2$.
The hyper parameters, i.e., the cooling factor $\eta$,  
the mini-batch size $K$,
and the number of mini-batches $t_{max}$,  
should be adjusted carefully to achieve  a satisfactory result.

\section{Experimental study: Gaussian source}

In order to study the fundamental behavior of the neural quantizer, we 
carried out numerical experiments for the simplest setting, where 
the source signal follows a Gaussian PDF.
Although the quantizer design problem for an i.i.d. Gaussian random variable 
can be solved efficiently using known algorithms, such as Lloyd algorithm \cite{Lloyd}, 
the problem is extremely simple and is therefore 
suitable for observing the behavior and properties
of the neural quantizer.

\subsection{Model}

The source signal $X$ follows the zero-mean Gaussian PDF with variance 1.
Here, we assume a transparent system, where $Y=X$ and $y = \D(y)$ hold.
Based on this assumption, the objective function to be minimized is 
$
	\E \left[||x^{(i)}  - q_{NQ}(x^{(i)}  ; \sigma^2, L))||_2^2 \right],
$
i.e., the expected squared $L_2$ distortion between the source signal and the corresponding 
quantized signals.

\subsection{Results}

\begin{figure}
\begin{center}
	\includegraphics[scale=0.7]{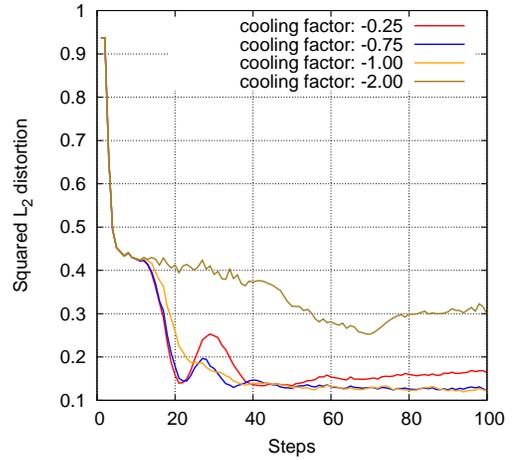}	
\end{center}
	\caption{Expected squared distortion as a function of training step ($L = 4$,  $u = 8$, $T=2$)}
	\label{gauss_exp1}
\end{figure}

We first discuss the choice of the cooling factor.
Figure \ref{gauss_exp1} presents the expected squared $L_2$ distortion as a function of training steps $t$
for the cases $\eta \in \{-0.25, -0.75, -1.0, -2.0 \}$.
The number of levels, the hidden state dimension, and the number of layers is
$L = 4$, $u = 8$, and  $T = 2$, respectively. 
In the training processes, the mini-batch size is set to $K = 100$.
From Fig. \ref{gauss_exp1}, it can be easily observed that
the expected squared $L_2$ distortion rapidly decreases 
as the number of training steps increases,  except in the case of $\eta = -2.0$.
This implies that the trainable variables in the neural quantizer are 
successfully updated to
lower the objective function in the supervised training processes.
The annealing process for $\eta = -2.0$ appears to be too fast to decrease the 
temperature parameter $\sigma^2$ and results in a higher floor of the objective function values.
For the cases of $\eta \in \{-0.25, -0.75, -1.0 \}$, 
the minimum distortion is achieved around step $t = 50$.
The cases of $\eta  = -0.75$ and $\eta  =-1.0$ provide the best result, 
i.e., the minimum saturated value after $t > 50$.
The observation obtained from this experiment is that an appropriate choice of
cooling factor is crucial.

\begin{figure}
\begin{center}
	\includegraphics[scale=0.7]{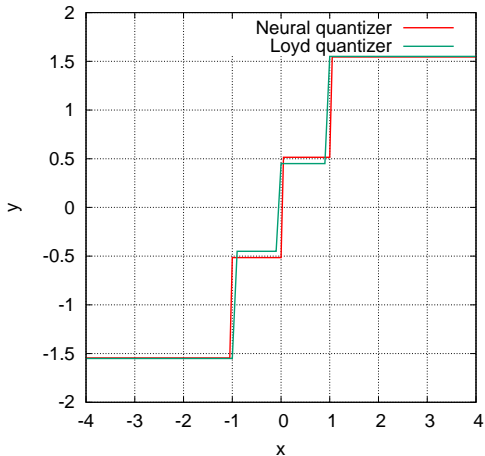}	
\end{center}
	\caption{A trained neural quantizer $y = q_{NQ}(x)$ for Gaussian source. 
	 ($L = 4$,  $u = 8, T= 2$, $\eta = -0.75, K = 100, t_{max} = 200$). As a benchmark, 	
	the quantization function of the Lloyd quantizer is also included.}
	\label{gauss_exp2}
\end{figure}

In this problem setting, Lloyd  algorithm \cite{Lloyd} 
can be used to find the optimal quantizer 
with respect to the expected squared $L_2$ distortion. 
The paper \cite{Lloyd} presents the optimal four-level quantizer 
obtained by Lloyd algorithm, referred to as {\em Lloyd quantizer}.
This Lloyd quantizer takes the values in  $\{-1.51, -0.45, 0.45, 1.51 \}$ 
for the above problem setup.  
The expected squared $L_2$ distortion of the Lloyd quantizer is  $0.12$.
Figure \ref{gauss_exp2} presents the output 
of a trained neural quantizer and Lloyd quantizer 
as a function of input.
The parameters of the neural quantizer are described in
the caption of Fig.\ref{gauss_exp2}. In this case, 
the neural quantizer takes the values of $\{-1.55, -0.52, 0.52, 1.55\}$,
which are fairly close to the quantization levels of the Lloyd quantizer.
The expected squared $L_2$ distortion of the neural quantizer is $0.12$ 
which approximately equals  
that of the optimal value obtained by the Lloyd algorithm.

\section{LDPC codes and sum-product algorithm}


\subsection{Model}

Let $C_H \subset \F_2^n$ be an LDPC code of length $n$,  
and let $H \in \F_2^{m \times n}$ be 
an LDPC matrix defining $C_H$.
We assume a standard LDPC-coded BPSK modulation scheme over an AWGN channel.
The details of the channel model are as follows.
Let $X$ be a random variable representing a transmitted codeword in $C_H$.
The function $\beta: \F_2 \to  \{+1, -1\}$ defined by
$
\beta(0) = 1, \quad \beta(1) = -1
$
is the binary-to-bipolar conversion function.
The random variable $Y$ representing a received word is given by
$
Y = \beta(X) + W.
$
The vector $W$ is  an AWGN vector, where 
each component of $W$ follows a zero-mean Gaussian 
distribution with variance $v^2$. 
The received word $Y$  is first quantized by the neural quantizer $q_{NQ}(\cdot)$
and the quantized signals are then passed to an LDPC decoder 
based on the log-domain sum-product algorithm \cite{Richardson}.

The log-domain sum-product algorithm is described in detail 
as follows. Let $G = (V, C, E)$  be the Tanner graph 
corresponding to the parity check matrix $H$. 
The set $V = [1,n]$ is the set of variable nodes and 
the set $C = [1, m]$ is the set of check nodes. 
The notation $[a, b]$ represents the set of
consecutive integers from $a$ to $b$.
The edge set is defined by
$
E \defeq \{  (i, j) \in C \times V \mid  H_{i,j} = 1 \}
$
according to the definition of the Tanner graph, 
where $H_{i,j}$ represents the $(i,j)$-component of $H$.
The set $A(i) (i \in [1,m])$ is defined by $A(i) \defeq \{j \in [1,n] \mid  H_{i, j} = 1 \}$ and
the set $B(j) (j \in [1, n])$ is defined by $B(j) \defeq \{i \in [1,m] \mid H_{i, j} = 1 \}$.

The variable node operation can be summarized as 
$
	\beta_{j \to i} = \lambda_j + \sum_{k \in B(j) \backslash i} \alpha_{k \to j},
$
where $\beta_{j \to i}$ is the message from variable node $j$ to check node $i$ 
and $\alpha_{k \to j}$ is the message from check node $k$ to variable node $j$.
The symbol $\lambda_j$ denotes the incoming log likelihood ratio (LLR)  calculated from the
received signal. For a realization of a received symbol $y_j$, the corresponding LLR value is $\lambda_j = 2 y_j /v^2$.
For each round, check node $i$ calculates 
the message from check node $i$ to variable node $j$, which is given by
$
	\alpha_{i \to j} = 2 \tanh^{-1} \left( \prod_{k \in A(i) \backslash j} \tanh \left( \frac 1 2 \beta_{k \to i}  \right)  \right).
$
The two operations, i.e., the variable and check node operations, 
are repeated to make the final estimation.

We here combine the neural quantizer and the log-domain sum-product 
algorithm in the following manner.
Let $y \in \R^n$ be a received word which is a realization of the random variable $Y$. 
The receiver 
first applies the neural quantizer to the received symbols,  and we obtain
$
	\lambda_j = q_{NQ}(y_j; \sigma^2, L),\ j \in [1, n].
$
These {\em quantized LLR values} are fed to the variable node operation for each iteration round.
We also apply the sigmoid function to the final output of the log-domain sum-product algorithm:
$
	\hat x_j = \sigma \left( \lambda_j + \sum_{k \in B(j)} \alpha_{k \to j} \right),
$
where $\sigma(\cdot)$ is the sigmoid function defined by $\sigma(x) \defeq (1 + \exp(-x))^{-1}$.
A threshold function is commonly used in the final round of the sum-product algorithm, but 
the function has zero gradient almost everywhere. 
Since we replaced the threshold function by the sigmoid function, a backpropagation algorithm 
can be used to 
calculate the gradients of the trainable variables in the neural quantizer.
Note that a backpropagation algorithm is applicable to the entire log-domain sum-product algorithm 
as shown by Nachmani et al. \cite{Com2}.
In the training processes, we used squared $L_2$ distortion 
as an objective function as in (\ref{L2dist}).

\subsection{Results}

\begin{figure}
\begin{center}
	\includegraphics[scale=0.7]{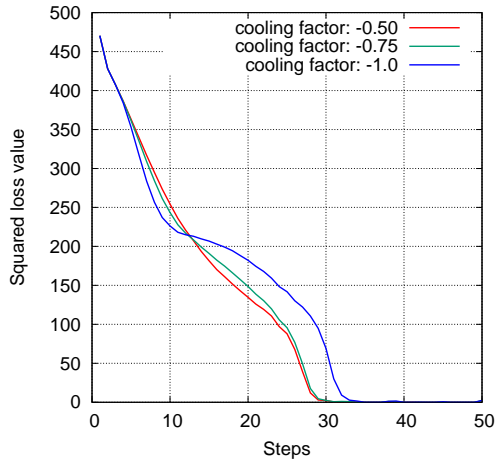}	
\end{center}
	\caption{Squared loss values as a function of training step ($L = 8$,  $u = 8, T= 2$)}
	\label{LDPC_exp1}
\end{figure}

Figure \ref{LDPC_exp1} shows the squared loss value as a function of training step.
In training processes,  the Adam optimizer with an initial value 0.04 was used.  
The LDPC code used  
in the experiments is a regular LDPC codes PEGReg504x1008 
\cite{PEG} $(n = 1008, m = 504)$ with 
a variable node degree of 3. 
We assume  an $8$-level neural quantizer with parameters $u = 8$ and $T= 2$.
During the training phase, the mini-batch size is set to $K = 100$ and 
the SNR of the channel is set to 2.5 dB for generating the samples.
In all of the experiments presented in this subsection,
the number of iterations in the sum-product decoder is fixed to 20.
Based on this figure, the cooling factor of $\eta = -1.0$ appears to be 
too small to achieve 
fast convergence.

\begin{figure}
\begin{center}
	\includegraphics[scale=0.7]{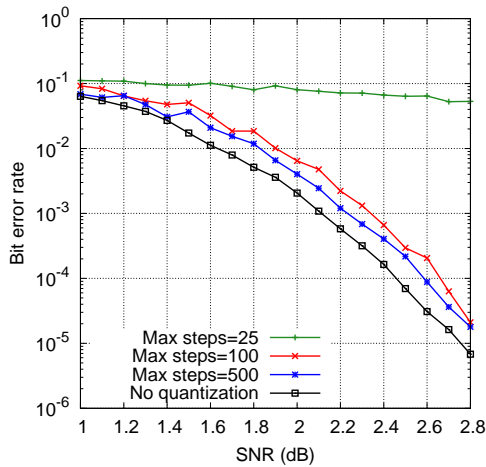}	
\end{center}
	\caption{BER curves of sum-product decoder with an 8-level trained neural quantizer ($L = 8$,  $u = 8, T= 2$). 
	LDPC code: PEGReg504x1008 $(n = 1008, m = 504)$.}
	\label{LDPC_exp3}
\end{figure}

Figure \ref{LDPC_exp3} presents the bit error rate (BER) curves 
of the quantized LDPC-coded BPSK scheme.
When measuring the BER performance after a training phase finished, 
we fixed $\sigma^2 = 0$.
This means that the neural quantizer becomes a hard quantizer.
We also used the hard threshold function instead of the sigmoid function at the final stage of
the sum-product decoder to generate $\hat x$.
We examined three cases in which the maximum number 
of training steps $t_{max}$ are 25, 100, and 500.
As a baseline, the BER curve of the sum-product decoder without quantization 
is included in Fig. \ref{LDPC_exp3} as well.
In the baseline system, 
the optimal LLR calculation rule $\lambda_j = 2 y_j / v^2 (j \in [1, n]$ is used.
In the training phase, we chose the cooling factor $\eta = -0.5$.
The BER performance of $t_{max} = 25$ is very poor 
because the quantized values were not appropriately learned.
In the experiments, the BER performance rapidly improves when $t_{max}  > 50$.  
The improvement is 
saturated around $t_{max} = 500$. The difference between the baseline BER curve and
the BER curve of $t_{max} = 500$ is approximately 0.1 dB over the entire range of SNR.
This result indicates 
that the joint optimization method is able to provide a quantizer with reasonable BER performance.

\begin{figure}
\begin{center}
	\includegraphics[scale=0.7]{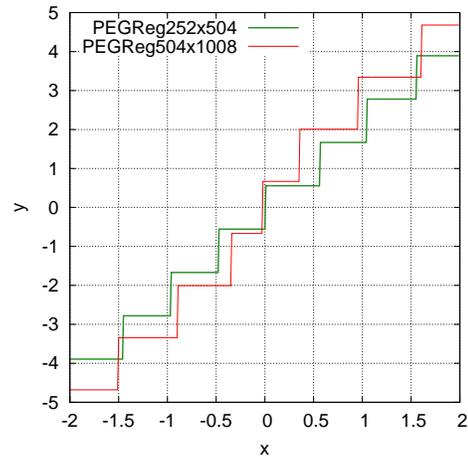}	
\end{center}
	\caption{Trained neural quantizers $y = q_{NQ}(x)$ 
	($L = 8$,  $u = 8, T= 2$), 
	LDPC codes: PEGReg504x1008 and PEGReg252x504.}
	\label{LDPC_exp2}
\end{figure}

Figure \ref{LDPC_exp2} plots the trained quantizer functions.
The experiment setting is the same as in the previous experiments 
dealing with PEGReg504x1008.
For the purpose of comparison, we included the result of a trained quantizer for a
regular LDPC code PEGReg252x504 \cite{PEG} with $n = 504$ and $m = 252$.
Figure \ref{LDPC_exp2} shows that an 8-level non-uniform quantizer function is learned 
in the training phase. Moreover, 
the trained quantizer automatically learned appropriate LLR conversion. 
Interestingly, the optimized quantizer for $n = 504$ is 
not the same as the one for $n = 1008$.
This may imply that the optimal quantizer is dependent on the choice of codes.

\section{Concluding summary}

In the present paper, a joint quantizer optimization method
for constructing a quantizer matched to the sum-product algorithms
was presented.
The neural quantizer allows us to exploit standard supervised 
learning techniques for DNNs to optimize the quantizer.
We limited our focus to the sum-product decoders, but 
the joint quantizer optimization method proposed herein is expected to be universal,
i.e., the proposed method can be applied to another decoding algorithm or
a detection algorithm if the algorithm can be described as a differentiable function.

\section*{Acknowledgment}
The present study  was supported in part by a JSPS Grant-in-Aid 
for Scientific Research (A) (Grant Number 17H01280),
and a JSPS Grant-in-Aid for Scientific Research (B) (Grant Number 16H02878).

\end{document}